# Electro-oxidation of a cobalt based steel in LiOH: a non-noble metal based electro-catalyst suitable for durable water-splitting in an acidic milieu

Helmut Schäfer,*[a] Karsten Küpper,[a,b] Klaus Müller-Buschbaum,[c] Diemo Daum,[d] Martin Steinhart,[a] Joachim Wollschläger,[a,b] Ulrich Krupp,[e] Mercedes Schmidt,[a] Weijia Han[a] and Johannes Stangl[c]

The use of proton exchange membrane (PEM) electrolyzers is the method of choice for the conversion of solar energy when frequently occurring changes of the current load are an issue. However, this technique requires electrolytes with low pH. All oxygen evolving electrodes working durably and actively in acids contain $IrO_x$. Due to their scarcity and high acquisition costs, noble elements like Pt, Ru and Ir need to be replaced by earth abundant elements. We have evaluated a cobalt containing steel for use as an oxygen-forming electrode in $H_2SO_4$. We found that the dissolving of ingredients out of the steel electrode at oxi-dative potential in sulfuric acid, which is a well-known, serious issue, can be substantially reduced when the steel is electro-oxidized in LiOH prior to electrocatalysis. Under optimized synthesis conditions a cobalt-containing tool steel was rendered into a durable oxygen evolution reaction (OER) electrocatalyst (weight loss: 39 μg $mm^{-2}$ after 50 000 s of chronopotentiometry at pH 1) that exhibits overpotentials down to 574 mV at 10 mA $cm^{-2}$ current density at pH 1. Focused ion beam SEM (FIB-SEM) was success-fully used to create a structure–stability relationship.

## 1. Introduction

Water can be converted into a fuel consisting of $H_2$ plus $O_2$ by the application of solar energy,[1–11] for instance, upon classical electrolysis. Due to the sluggish kinetics of one of the half-cell reactions, the water oxidation reaction, the electrochemical cleavage of water molecules cannot be performed at the theor-etically required (thermodynamic) potential and the oxygen evolution reaction on common electrode materials is accompanied by high overpotentials.[12,13] The efficiency and stability of the anodes used for water electrolysis critically depend on the pH value of the electrolyte. Currently, electrocatalysts, even when based on non-noble metals, are known to be efficient and durable toward OER at high pH values.[1,14–19] However, alkaline electrolyzers are less resistant against fre-quently occurring changes of the current load.[1] Proton exchange membrane (PEM) electrolyzers not only benefit from higher gas purity and higher efficiency, but also are suitable for the storage of renewable energy which is basically charac-terized by high dynamics.[1] But the operation of these electro-lyzers requires a low pH condition. Ir–Ru-oxides are among the known materials that are considered to be suitable electrocata-lysts in the acidic regime, exhibiting both high activity and reasonable durability. In particular, the stability of non-noble-metal-based electrocatalysts towards oxidative water-splitting of acids needs to be optimized.[20–27] Therefore, the develop-ment of electrocatalysts solely based on nonprecious metals suitable for robust and efficient anodic water-splitting in the acidic regime is of the highest interest,[22] and certainly rep-resents the topic of ongoing research in many groups.

Surface modified steels are known to be efficient, stable, cheap and easily accessible electrode materials ideally suited for the electrochemically driven cleavage of water into its elements.[20,28–31] Until recently, we failed to render steel into an electrocatalyst that can be considered as competitive to Ir–Ru-containing OER electrocatalysts for water-splitting in acids.[20] In this report we evaluate the suitability of a Co based steel as an OER electrocatalyst for the water splitting reaction performed in the acidic regime. X20CoCrWMo10-9 steel, basi-cally consisting of Fe, Co, Cr, Mo and W, electro-oxidized in LiOH showed reasonable performance and stability towards long term usage as an OER electrocatalyst in 0.05 M $H_2SO_4$ and was found to be highly competitive to recently developed and significantly more expensive OER electrocatalysts working in the acidic environment.

[a]Institute of Chemistry of New Materials and Center of Physics and Chemistry of New Materials, Universität Osnabrück, Barbarastrasse 7, 49076 Osnabrück, Germany. E-mail: helmut.schaefer@uos.de
[b]Department of Physics, Universität Osnabrück, Barbarastraße 7, 49069 Osnabrück, Germany
[c]University of Würzburg, Institute of Inorganic Chemistry, Julius-Maximilians-Universität Würzburg, Am Hubland, D-97074 Würzburg, Germany
[d]Faculty of Agricultural Science and Landscape Architecture, Laboratory of Plant Nutrition and Chemistry, Osnabrück University of Applied Sciences, Am Krümpel 31, 49090 Osnabrück, Germany
[e]Institute of Materials and Structural Integrity, University of Applied Sciences Osnabrueck, Albrechtstraße 30, 49076 Osnabrück, Germany

# 2. Experimental section

2.1 Sample preparation

2.1.1 Samples made from untreated steels (sample Co). Samples with a total geometry of 70 × 10 × 1.5 mm were constructed from 1.5 mm thick sheets consisting of X20CoCrWMo10-9 steel. X20CoCrWMo10-9 steel was pur-chased from WST Werkzeug Stahl Center GmbH & Co. KG, D-90587 Veitsbronn-Siegelsdorf, Germany. Pre-treatment: the surface of the metal was cleaned intensively with ethanol and polished with grit 240 SiC sanding paper. Afterwards, the surface was rinsed intensively with deionized water and dried under air for 100 min at room temperature.

2.1.2 Sample Co-Cy. The oxide layer of sample Co-Cy was grown using a repetitive potential multicycling technique within a conventional three electrode set-up consisting of a metal WE (sample Co), a platinum wire CE (5 × 4 cm geometric area) and a reversible hydrogen reference electrode (HydroFlex, Gaskatel Gesellschaft für Gassysteme durch Katalyse und Elektrochemie mbH, D-34127 Kassel, Germany). An Interface 1000 potentiostat from Gamry Instruments (Warminster, PA 18974, USA) was employed and interfaced to a personal com-puter which allowed recording all the electrochemical data digitally. The WE (anode) was immersed 1.5 cm deep (3.4 $cm^2$ geometric area), and the CE (cathode) was completely immersed in the electrolyte which was prepared as follows: in a 250 mL glass beaker, 16 g (95 mmol) of $LiOH·8H_2O$ (VWR, Darmstadt) was dissolved under stirring and cooling in 140 g deionized water. The anodization was performed under stir-ring (300 rpm) using a magnetic stirrer (20 mm stirring bar). The RE was placed between the working electrode and the CE. The distance between the WE and the RE and the distance between the RE and the CE were adjusted to 4–5 mm. The potential of the WE vs. RHE was varied between −0.1 and +1.65 V. The sweep rate was set at 10 mV $s^{-1}$ and the step size was 20 mV. The experiment was completed after 3000 cycles which corresponds to the total duration of the experiment of 291.6 h. It turned out that the current peaks are still below the current limit of the device (1000 mA). In order to check the reproducibility, the experiment was repeated four times. Before and after carrying out the electro-oxidation procedure the weight of the specimen had been determined using a precise balance (Sartorius 1712, 0.01 mg accuracy; Table S3†). After completing the experiment the CE and the WE were taken out of the electrolyte and rinsed intensively with tap water for 15 min and afterwards with deionized water for a further 10 min. Prior to the electrochemical characterization the samples were dried under air at ambient temperature.

All other sample series have been prepared according to the procedure described in the ESI.†

2.2 Electrochemical measurements

A three-electrode set-up was used for all electrochemical measurements. An apparent surface area of 2 $cm^2$ was defined on the working electrode (WE) by an insulating tape (Kapton tape). The Ir-$RuO_2$ sample (10 micrometer layer deposited on titanium) with a total geometry of 100 × 100 × 1.5 mm was pur-chased from Baoji Changli Special Metal Co, Baoji, China. An electrode area of 2 $cm^2$ was defined on the plate by Kapton tape. To avoid additional contact resistance the plate was electrically connected via a screw. A platinum wire electrode (4 × 5 cm geometric area) was employed as the CE, and a reversible hydrogen reference electrode (RHE, HydroFlex, Gaskatel Gesellschaft für Gassysteme durch Katalyse und Elektrochemie mbH, D-34127 Kassel, Germany) was utilized as the reference standard; therefore all voltages are quoted against this refer-ence electrode (RE). For all measurements the RE was placed between the working electrode and the CE. The measurements were performed in a 0.05 M $H_2SO_4$ (VWR, Darmstadt, Germany) solution, respectively. Measurements were per-formed at room temperature (295.15 K). The distance between the WE and the RE was adjusted to 1 mm and the distance between the RE and the CE was adjusted to 4–5 mm. Voltage drop compensation was realized by 60% compensation of the solution resistance shown in Table 1 determined by frequency response analysis measurements. The corrected voltages were denoted as E-IR. All electrochemical data were recorded digi-tally using an Interface 1000 potentiostat from Gamry Instruments (Warminster, PA 18974, USA), which was inter-faced to a personal computer.

Cyclic Voltammograms (CV) were recorded in 90 mL of electro-lyte in a 100 mL glass beaker under stirring (450 rpm) using a magnetic stirrer (21 mm stirring bar). The scan rate was set at 20 mV $s^{-1}$ and the step size was 2 mV. The potential was cyclically varied between 1.2 and 1.9 V vs. RHE for OER measurements.

Chronopotentiometry scans were conducted at a constant current density of 10 mA $cm^{-2}$ in 90 mL of electrolyte for measur-ing periods <2000 s, and in 800 mL of electrolyte for measuring periods ≥10 000 s, respectively. The scans were recorded under stirring (450 rpm) using a magnetic stirrer (25 mm stirring bar) for measuring periods <2000 s, and using a magnetic stirrer (40 mm stirring bar) for measuring periods ≥10 000 s, respect-ively. Before and after carrying out long term chronopotentio-metry measurements the weight of the specimen was determined by using a precise balance (Sartorius 1712, 0.01 mg accuracy).

Table 1 Overview of the prepared samples (columns I), the performed surface modification (columns II and III) as well as the electrocatalytic pro-perties of the samples (columns IV–VII); standard errors in square brackets

| Sample name/material | Activation Electro-ox. | Activation Therm. | $J_{max}$ at 1.9 V vs. RHE derived from CV | Average potential (V vs. RHE) for 10 mA cm$^{-2}$ | Average weight loss (μg mm$^{-2}$) after 50 000 s of chronopotentiometry at 10 mA cm$^{-2}$ | Resistivity $R_s/R_{CT}$ (Ω) at offset potential (V vs. RHE) derived from EIS |
|---|---|---|---|---|---|---|
| Co/X20CoCrWMo10-9 | — | — | 11 [0.4] | 1.924 [0.02] | 98.9 [2.02] | 5.8 [0.2]/14.5 [0.4] (1.8); 6.1 [0.2]/3 [0.2] (1.9) |
| Co-Cy/X20CoCrWMo10-9 | 0.68 M LiOH 3000 cycles | — | 9.3 [0.2] | 1.802 [0.02] | 39.1 [1.21] | 5 [0.3]/14 [0.6] (1.7); 4.4 [0.3]/3 [0.3] (1.9) |
| Co-300.1/X20CoCrWMo10-9 | 4.8 M NaOH/ 300 min | — | 8.0 [0.3] | 1.875 [0.03] | 94.9 [2.22] | |
| IrO$_2$–RuO$_2$ | — | — | 32 [0.8] | 1.710 [0.04] | 8 [0.2] | |

## 3. Results and discussion

### 3.1 OER properties of untreated stainless steels

We have examined three different steel samples as potential OER electrocatalysts in $H_2SO_4$. This includes untreated cobalt steel as well as surface modified cobalt steels. Table 1 gives an overview of the samples/sample preparations and the corresponding OER key data. Five representatives of each sample have been synthesized and investigated. Fig. 1 summarizes the electrocatalytic properties of the untreated steel sample for the OER in 0.05 M $H_2SO_4$. As expected, the specimen consisting of untreated stainless steel X20CoCrWMo10-9- steel (sample Co) showed a significant increment of the current at around 1.2 V vs. RHE derived from the CV measurements due to oxidation of the catalyst itself (Fig. 1a). Onset of oxygen evolution (sample Co; Fig. 1a) takes place at around 1.77 V vs. RHE. The sample exhibited a reason-able and stable potential at a constant current density under steady state conditions (sample Co; Fig. 1a). Simultaneously intensive bubble formation upon the surface of sample Co can be clearly seen. The overpotential amounted to 696 mV (sample Co) at 10 mA cm$^{-2}$ current density in 0.05 M $H_2SO_4$ (Fig. 1a). However, the OER performance of sample Co was found to be sensitive towards repeated dynamic variation of the voltage (Fig. 1b) between 1.2 and 1.9 V vs. RHE: thus e.g. the current density reached at 1.85 V vs. RHE decreased from 6.29 to 4.41 mA cm$^{-2}$ after 1000 cycles (sample Co, Fig. 1b), respectively. In addition to a weakening OER-based current density to voltage ratio, the dissolving of steel ingredients out of the electrode at oxidative potentials turned out to be a serious issue. Untreated steel X20CoCrWMo10-9 lost on average 98.9 μg mm$^{-2}$ upon OER polarization for 50 000 s in 0.05 M $H_2SO_4$ at 10 mA cm$^{-2}$ (Table 1). For practical applications, this is by far too much. Thus, for instance, under identical conditions the weight loss of IrO$_2$–RuO$_2$ was less than one tenth of this value (8 μg mm$^{-2}$; Table 1). In order to verify this mass loss of sample Co during OER, we double checked the release of steel ingredients out of the electrode into the electrolyte by perform-ing an ICP-OES analysis of the electrolyte used for long term chronopotentiometry. The elements determined via ICP-OES (Table S1†) after 50 000 s of OER in the electrolyte (0.05 M $H_2SO_4$) are reasonable in light of the steel composition.[29,32] The total amount of ions determined in the electrolyte is in good agreement with the mass deficit that occurred in the samples during long term OER electrocatalysis (columns II and V of Table S1†). X20CoCrWMo10-9 released Fe, Cr and Co besides W and Mo whilst chronopotentiometry was carried out in 0.05 M $H_2SO_4$.

Impedance spectroscopic investigations have been per-formed with sample Co at pH 1. We found that modeling of the frequency response behavior (0.1 Hz–50 kHz) of all samples discussed in this work at an offset potential that ensures that oxygen evolution can be done using the circuit of the so-called Randles cell (Fig. 2a) in which the double layer capacity is in parallel with the impedance due to the charge transfer reaction. Therefore the Nyquist plot always shows a semicircle. The real axis value at the high frequency intercept can be interpreted as the solution resistance ($R_s$), and the real axis value at the low frequency intercept can be interpreted as the sum of the solution ($R_s$) and the charge transfer (CT) resist-ances ($R_{ct}$), respectively. Hence the diameter of the semicircle is equal to the CT resistance. The values for CT and solution resistances ($R_s$) of all samples can be taken from Table 1, and from Fig. 2b, respectively. At a significant overpotential of 672 mV (1.9 V vs. RHE) X20CoCrWMo10-9 steel exhibited a charge transfer resistance of 3 Ω (Fig. 2b, Table 1) that was found to be substantially increased to 14.5 Ω at 1.8 V vs. RHE, which corresponds to 572 mV overpotential. This can be easily explained: generally, the radii of the corresponding circle in the Nyquist plot for one and the same sample decrease with increasing offset potential, which originates from an accelera-tion of the charge transfer (Fig. 2b).

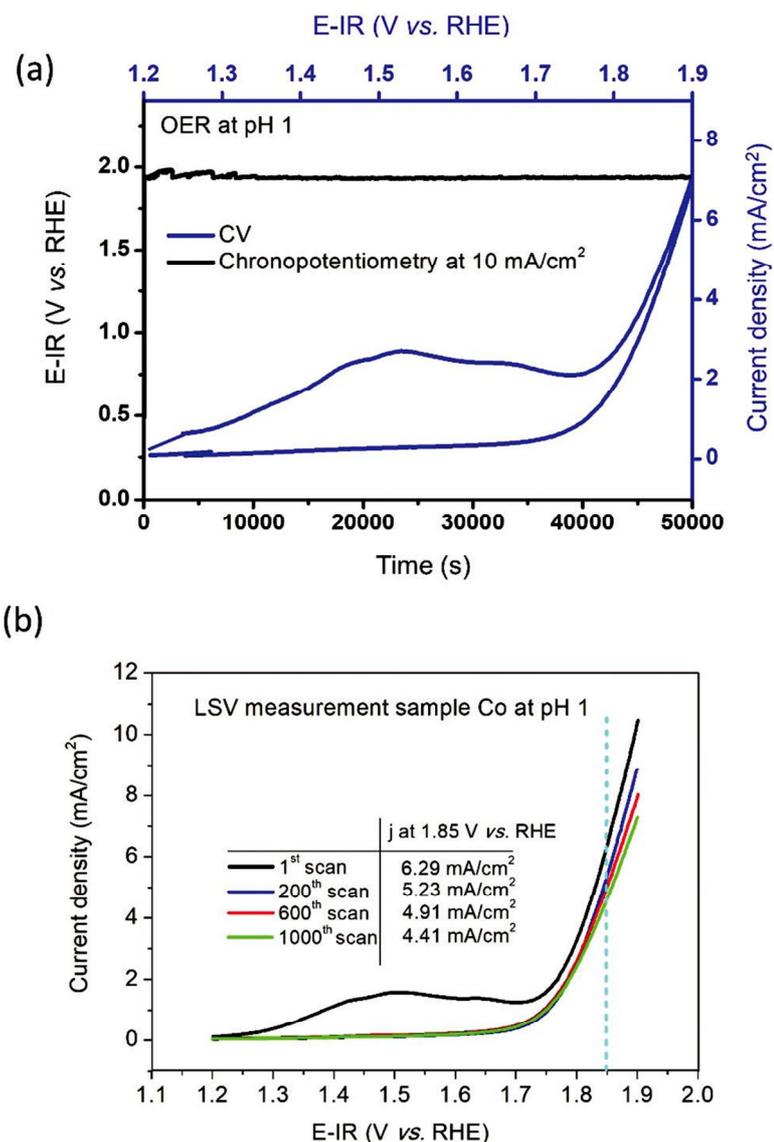

Fig. 1 Steady state and non-steady state voltage current behavior of sample Co in 0.05 M $H_2SO_4$. CVs were recorded with a scan rate of 20 mV $s^{-1}$. Linear sweep voltmetry (LSV) was performed with a scan rate of 10 mV $s^{-1}$. Electrode area of all samples: 2 $cm^2$. Stirring of the electrolyte was performed for all measurements. (a) Cyclic voltammogram of sample Co; long term chronopotentiometry plot of sample Co at a current density of 10 mA $cm^{-2}$. (b) LSV measurements performed with sample Co at 1.85 V vs. RHE.

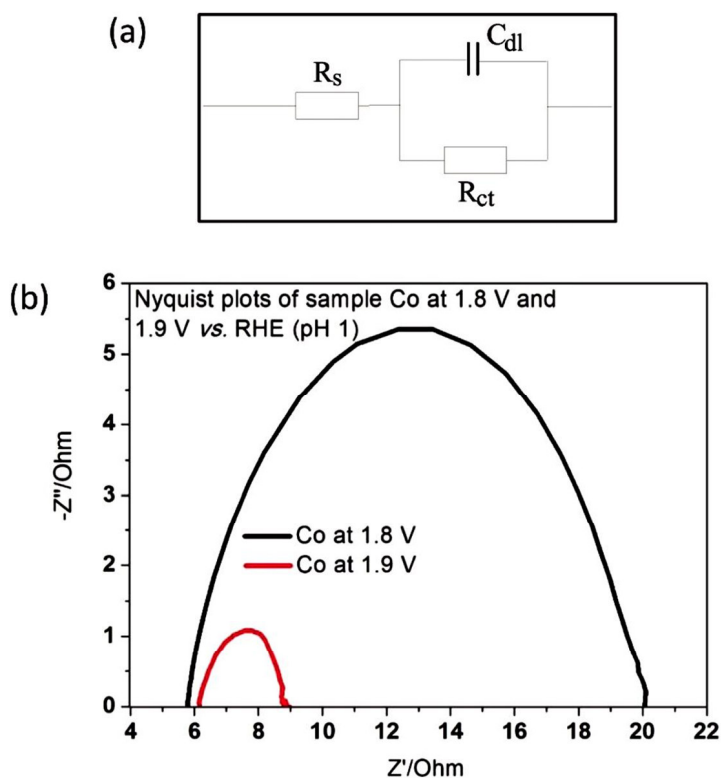

Fig. 2 (a) The circuit of the Randles cell. (b) Nyquiest plots of the frequency response analysis of samples Co at pH 1 and at offset potentials of 1.8 and 1.9 V vs. RHE.

### 3.2 OER properties of surface modified steels

The surface modification procedures applied to X20CoCrWMo10-9 steel basically aim at an improvement of the stability of the Co-containing steel toward OER in the acidic regime. The outcome of earlier studies dealing with the exploitation of steel and surface modified steel as OER electro-des in alkaline and neutral regimes, however, raised hopes for an improvement of the overall OER performance of X20CoCrWMo10-9 steel towards electrocatalytically initiated OER in an acidic environment upon suitable surface modifi-cation. As described in our previous publication, the electro-oxidation of X20CoCrWMo10-9 steel in NaOH turned out to be highly efficient in improving the electrocatalytic properties of the material towards OER in neutral media.[29] However, a transfer of the sample preparation technique established in our previous paper to this new study did not lead to a satisfying outcome. We applied an electro-oxidation procedure to the Co-steel carried out in 4.8 M NaOH for 300 min (Table 1) leading to a sample henceforth denoted Co.300.1. The current–voltage relationship in pH 1 electrolyte was positively influenced in the sense that the potential derived from chronopotentiometry measurements of Co-300.1 ensuring a current density of 10 mA cm$^{-2}$ (1.875 V vs. RHE; Fig. 3) was slightly reduced when compared to the corres-ponding value of sample Co (1.924 V vs. RHE, Fig. 1a). In addition, the current density derived from the CV of sample Co-300.1 substantially faster increased than the one in the CV of sample Co.

As a matter of fact, sample Co-300.1 is at the level of recently developed OER electrocatalysts like $Ba_2TbIrO_6$,[26] $MnCoTaSbO_x$[33] and electrodes with 10 wt% $Ir_{0.5}Ru_{0.5}O_2$.[48] However, we could not overcome the main drawback of sample Co: the significant weight loss during long term OER polarization (sample Co: 98.9 μg mm$^{-2}$; sample Co-300.1: 94.9 μg mm$^{-2}$) was not reduced noticeably (Table 1). It turned out that the compositions of the electrolytes used for long term chronopotentiometry of samples Co and Co-300.1 are similar and just the absolute ion content is higher for the electrolytes used for sample Co (Table S1†). The total amount of ions determined in the electrolyte is in good agreement with the mass deficit that occurred in sample Co-300.1 during long term OER electrocatalysis (Table S1†). We modified the electro-oxidation procedure and used henceforward LiOH instead of NaOH. Samples made of X20CoCrWMo10-9 steel were electro-oxidized in 0.68 M LiOH upon cycling oxidation (samples Co-Cy). Sample Co-Cy presents the best outcome of this study exhibiting a mass loss of 39.1 μg mm$^{-2}$ after 50 000 s of chron-opotentiometry at 10 mA cm$^{-2}$ (Table 1) at pH 1, which represents a reduction when compared to untreated steel by ~60% (sample Co = 98.9 μg mm$^{-2}$). Again the mass loss was reasonably verified by an ICP-OES study carried out with the electrolyte. Lithium was not determined in the electrolyte (Table S1†). Degradation of electrocatalysts in acids is well known[34,35,43] after long term polarization at positive potentials.

MnO$_x$ has been exploited as an OER electrocatalyst in different acidic regimes with pH values in between −0.5 and 2.[36] However, current densities were found to be extremely low, ≪1 mA cm$^{-2}$ (Table 2), at reasonable potentials and significant dissolution was found. Layered manganese–calcium oxide was investigated as a prospective OER electrocatalyst in 0.1 M HClO$_4$ but dissolved even without oxidative potentials and exhibited a poor OER efficiency.[37]

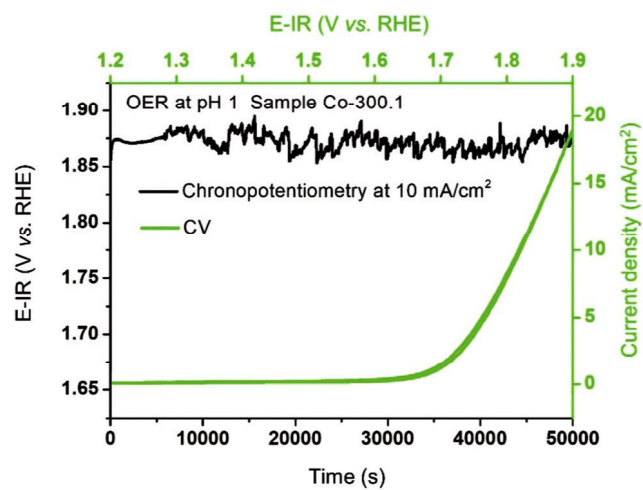

Fig. 3 The OER properties of surface oxidized stainless steel X20CoCrWMo10-9 in 0.05 M H$_2$SO$_4$ investigated under non-steady state and steady state conditions. The CV was recorded with a scan rate of 20 mV s$^{-1}$. Chronopotentiometry measurements were performed at mA cm$^{-2}$ current density. The electrode area of the sample: 2 cm$^2$. Stirring was applied to the electrolyte for all measurements. Cyclic voltammogram and chronopotentiometry plot of sample Co-300.1. Average overpotential through the 50 000 s scan: 647 mV.

A low overpotential ($\eta$ = 574 mV) determined at 10 mA cm$^{-2}$ current density can be derived from the chronopotentiometry plot (Fig. 4a) and makes Co-Cy a highly competitive OER electrocatalyst at pH 1. The non-steady state OER performance of sample Co-Cy ($j \sim 21$ mA cm$^{-2}$) was significantly better than that of sample Co ($\sim 7$ mA cm$^{-2}$ at 1.9 V vs. RHE) (Fig. 4a and 1a). The onset of OER in 0.05 M H$_2$SO$_4$ started at a potential as low as 1.70 V vs. RHE (Fig. 4a). CoTiP was recently studied[38] in 0.5 M H$_2$SO$_4$ and was inferior ($\eta$ = 971 mV at 8 mA cm$^{-2}$) to sample Co-Cy. In addition, non-noble element containing compounds like CoP,[22] NiB,[22] CoB,[22] NiMoFe,[22] NiFe,[22] CoO$_x$,[21] NiFeO$_x$,[21] and NiO$_x$[21] were found to be extremely unstable towards OER in 1 M H$_2$SO$_4$ and less active with $\eta \sim 1000$ mV at 10 mA cm$^{-2}$ (ref. 22). Frydendal reported on Ti stabilized MnO$_2$ as a prospective OER electrocatalyst and determined that $\eta \sim 550$ mV at 3 mA cm$^{-2}$ in 0.05 M H$_2$SO$_4$.[39] Very recently, Pi et al. reported on overall water splitting at low pH value upon IrM (M = Ni, Co, Fe) nanoparticles and determined for the IrNi compound an overpotential of ∼300 mV at 5 mA cm$^{-2}$ current density in 0.1 M HClO$_4$ derived from galvanostatic measure-ments.[40] Electrodeposited CoFePbO$_x$ films were recently inves-tigated as an OER electrocatalyst at pH 2.5 and were found to be stable towards 12 h of chronopotentiometry carried out at 1 mA cm$^{-2}$ (ref. 41). An F-doped CuMn-oxide based OER- and oxygen reduction electrode intended to be suitable for electro-catalysis in sulfuric acid was recently shown.[42] Unfortunately, a detailed evaluation of the mass loss during usage has not been shown. In addition, the OER activity, when derived from chronopotentiometry data, was mediocre ($\eta$ = 320 mV at ∼1.5 mA cm$^{-2}$ in 0.5 M H$_2$SO$_4$). The best OER performance achieved in acids upon a noble metal containing catalyst (IrO$_x$//SrIrO$_3$) was shown by Seitz et al. ($\eta$ = 280 mV at 10 mA cm$^{-2}$).[43] Notably, the overpotential for the OER was deter-mined in 0.5 M sulfuric acid, i.e. the acid concentration was ten times higher than in our case (0.05 M H$_2$SO$_4$). However, the material was not 100% stable at oxidative potentials: Sr was determined via ICP-OES in the electrolyte used for long term polarization experiments.[43] Repeated LSV is a common tool to simulate fast aging of the electrode. Co-Cy exhibited significantly higher stability than Co towards OER at pH 1 based on repeated LSV scans (Fig. 1b and 4b). The dynamic potential–current behavior of Co-Cy did not substantially change under repeated execution (Fig. 4b). Under identical conditions (1.85 V vs. RHE), the current density drop after 1000 scans was reduced from 1.88 mA cm$^{-2}$ (Co, Fig. 1b) down to 0.05 mA cm$^{-2}$ (Co-Cy, Fig. 4b).

Table 2. The OER key data of recently developed electrocatalysts determined at low pH

| Sample | Average overpotential (mV) at 10 mA cm$^{-2}$ (pH) | Tafel slope (pH) | Faradaic efficiency (mA cm$^{-2}$) | Average weight loss (μg mm$^{-2}$) after 50 000 s of chronopotentiometry in 0.05 M H$_2$SO$_4$ at 10 mA cm$^{-2}$ | Ref. |
|---|---|---|---|---|---|
| Surface modified steel (Co-Cy) | 574 (1) | | 95.2% (10 mA cm$^{-2}$) | 39.1 | This work |
| IrO$_x$/SrIrO$_3$ | 280 (0) | ~39 mV dec$^{-1}$ | Was not quantified | — | 43 |
| Cu$_{1.5}$Mn$_{1.5}$O$_4$ | 322 (0) at 1.5 mA cm$^{-2}$ | 65.7 mV dec$^{-1}$ | ~100% (1.5 mA cm$^{-2}$) | — | 42 |
| CoO$_x$ | ~400 (3.4) at 1 mA cm$^{-2}$ | 529 mV dec$^{-1}$ (0) | ~95% at pH 3.4 | — | 21 |
| RuO$_2$–IrO$_2$ (Ir–RuO$_2$) PVD sputtered on titania | 482 (1) | | — | 8 | This work |
| MnO$_x$ | 219 (1) at 1 μA cm$^{-2}$ | 127 mV dec$^{-1}$ (2.5) | ~100% at pH 2.5 (0.3 mA cm$^{-2}$) | — | 36 |
| CoTiP | 972 (0) at 8 mA cm$^{-2}$ | — | — | — | 38 |
| Ti-Stabilized MnO$_2$ | 547 (1) at 3 mA cm$^{-2}$ | 170 mV dec$^{-1}$ (1) | — | — | 39 |

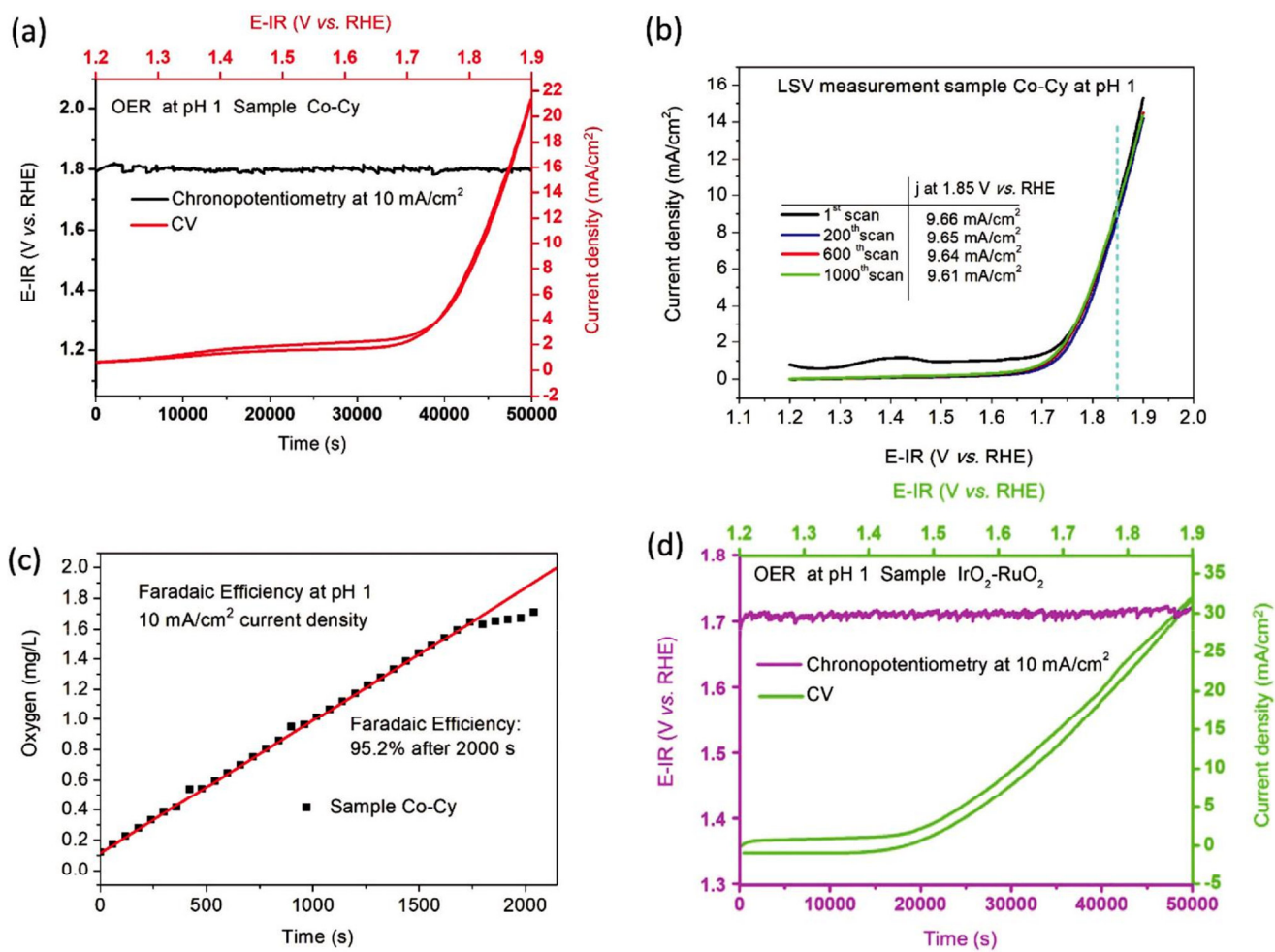

Fig. 4 The OER properties of samples Co-Cy and IrO$_2$–RuO$_2$ in 0.05 M H$_2$SO$_4$. CVs were recorded with a scan rate of 20 mV s$^{-1}$. The chronopoten-tiometry measurements were performed at 10 mA cm$^{-2}$ current density. Electrode area of all samples: 2 cm$^2$. Stirring was applied to the electrolyte for all measurements. (a) Cyclic voltammogram and chronopotentiometry plot of sample Co-Cy. (b) LSV measurements performed with sample Co-Cy at a potential of 1.85 V vs. RHE. (c) Correlation of oxygen evolution upon sample Co-Cy in 0.05 M H$_2$SO$_4$ (dotted curve) with the charge passed through the electrode system (the red line corresponds to 100% Faradaic efficiency). Electrode area of the sample: 2 cm$^2$; top: Faradaic efficiency of sample Co-Cy at 10 mA cm$^{-2}$ current density; total volume = 1890 mL, Faradaic efficiency after 2000 s: 95.2%; line equation: y = 0.000877368 × x + 0.12, where y represents oxygen content (mg L$^{-1}$) and x represents time (s). (d) Cyclic voltammogram and chronopotentiometry plot of the sample IrO$_2$–RuO$_2$.

However, to further exclude also "inner oxidation" (oxidation of the metal matrix below the oxide layer) during operation, it is indispensable to quantify the real oxygen evol-ution efficiency. The Faradaic efficiency for the OER upon sample Co-Cy at pH 1 at 10 mA cm$^{-2}$ amounted to 92.5% after 2000 s running time (Fig. 4c). These are reasonable efficiencies for the OER upon a non-noble metal based electrocatalyst in highly corrosive media. Anodized AISI 304 steel exhibited in 0.1 M KOH at 10 mA cm$^{-2}$ 75.5% charge-to-oxygen conversion[28] after 4000 s. Noble metal containing catalysts,[44,45] especially $IrO_2$–$RuO_2$,[21,46,47] are known for their high OER efficiency in the acidic regime. We have chosen commercially available $IrO_2$–$RuO_2$ sputtered on titanium as the reference sample (sample $IrO_2$–$RuO_2$) for OER activity and stability at pH 1. The onset of OER can be derived from Fig. 4d and amounted to ~1.45 V vs. RHE and agrees very well with the data from the literature.[27,48] As a matter of fact, the reference compound $IrO_2$–$RuO_2$ still exhibited OER properties superior to the ones of Co-Cy (Fig. 4a and d). The potential required to ensure 10 mA cm$^{-2}$ current density was 1.71 V vs. RHE, which is ~90 mV below the corres-ponding value of sample Co-Cy. Also, the non-steady-state current voltage behaviour was found to be slightly stronger than that of sample Co-Cy ($IrO_2$–$RuO_2$ = 32 mA cm$^{-2}$ at 1.9 V vs. RHE; Fig. 4d). However, the advantage of $IrO_2$–$RuO_2$ over sample Co-Cy with respect to current voltage behavior was not as substantial as expected and, moreover, even $RuO_2$-$IrO_2$ shows a "bleeding effect" (8 μg mm$^{-2}$; Table 1) when used as an OER electrode in the acidic regime.[23,49] Frequency response analysis carried out with samples Co and Co-Cy (Fig. 5) are in agreement with the results derived from DC-polarization experiments (Fig. 4). In the case that DC currents are applied, the resistance rises to infinity and the total resistance is represented by the sum of solution and charge transfer resistances (a simple Randles cell). The total resistance of sample Co-Cy occurring in diluted sulfuric acid derived from an EIS investigation performed at 1.9 V vs. RHE (7.4 Ω) was found to be significantly lower than the one deter-mined for sample Co (9.1 Ω), which reasonably explains the lower overpotentials for the OER upon the surface of sample Co-Cy in DC-polarization experiments.

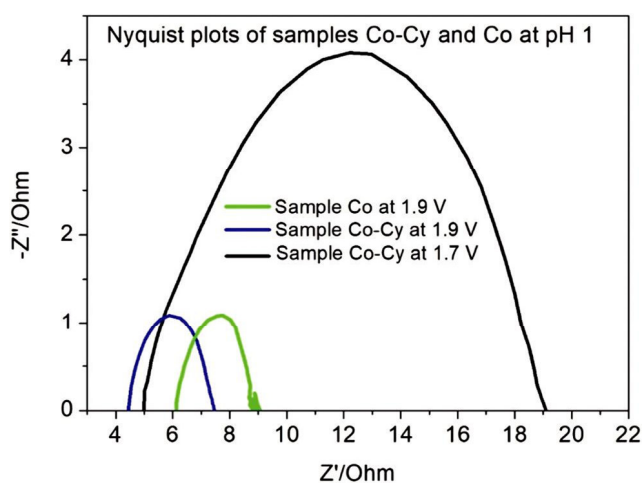

Fig. 5 Nyquiest plots of samples Co-Cy and Co. The offset potential was set at 1.7 and 1.9 V vs. RHE.

3.2.1 The origin of the improvement of the OER properties of X20CoCrWMo10-9 steel upon surface modification with LiOH. Multiple point nitrogen gas adsorption BET measurements (Fig. S1†) were carried out in order to determine the change in specific surface area through electro-oxidation. We showed in our previous publications that electro-oxidation of stainless steels Ni42 and X20CoCrWMo10-9 in 7.2 M NaOH did not lead to substantial changes in the size of the specific surface area.[20,29] However, electro-oxidation of a steel specimen consisting of X20CoCrWMo10-9 steel in diluted LiOH according to the protocol given in the experimental part was found to be accompanied by a substantial increment of the surface roughness. Thus sample Co-Cy indeed showed a higher specific surface area (0.567 m$^2$ g$^{-1}$) than sample Co (0.353 m$^2$ g$^{-1}$ (ref. 29)). This increment in roughness when compared to sample Co due to electro-oxidation can also be derived from SEM experiments (Fig. 6a–d) exhibiting holes and small cracks in the surface of sample Co-Cy. Notably, current densities given in this work are based on the projected area of the electrode. Therefore this increment in roughness certainly contributes to the substantially stronger current density–voltage ratio of samples Co-Cy when compared with sample Co as seen in Fig. 1a, 4a and summarized in Table 1. The XRD diffractogram of sample Co-Cy, recorded in grazing incidence mode (Fig. S2†), does not indicate crystalline oxides in the surface layer and the appearance of the periphery as seen by SEM experiments is rather typical for non-crystalline solids (Fig. 6c and d). A backscattered electron image of the microstructure of Co and Co-Cy obtained from a focused ion beam SEM study is shown in Fig. 7a and b. A ferrite phase with carbides well known from Co-containing tool steels[50] can be seen in both cases; however, the size of the carbides located in the periphery was found to be substantially reduced after surface modifi-cation (Fig. 7b). It is reasonable to assume that this is the origin of the smaller charge transfer resistance ($R_{CT}$) of sample Co-Cy (14 Ω) determined at an offset potential of 1.7 V vs. RHE (Fig. 5, Table 1) as compared to $R_{CT}$ of sample Co (14.5 Ω) determined at 1.8 V vs. RHE (Fig. 2b, Table 1). A lower charge transfer resistance will result in a lower voltage drop across the catalytically active outer oxide zone during electrocatalytically initiated oxygen evolution reaction and finally contribute to a lower OER based overpotential. Regarding the origin of the reduction of the ferritic carbides we can only speculate. Notably, this electro activation procedure goes hand in hand with a change of the surface composition (Table S2†). However, no mass loss occurred during the electro-oxidation of X20CoCrWMo10-9 steel upon repeated cycling of the poten-tial in diluted LiOH (Table S3†) which excludes a dissolution of some of the ingredients as the origin for the changes within the surface. This therefore suggests that electron migration takes place during the electro activation of the steel and is responsible for changes of surface composition and the reduction of the size of the ferritic carbides. Moreover, sample Co-Cy does not exhibit a classical sub-state-layer architecture known for samples achieved from electro deposition techniques. As shown in our previous report[29] this can be additionally seen as a source of stability during electrocatalytically initiated long term OER. The importance of the interface between the active catalyst and the substrate was recently proved by Yang et al. for carbon coated $Co_3O_4$ nanoarrays exploited as an oxygen evolution cata-lyst in acidic media.[51] A detailed investigation of the chemical nature of the surface of sample Co-Cy has been realized by XPS spec-troscopy. Cobalt was found to be completely suppressed and a Fe–Cr oxide containing outer sphere (89% Fe, 9.78% Cr; Table S2†) was created during the surface oxidation process.

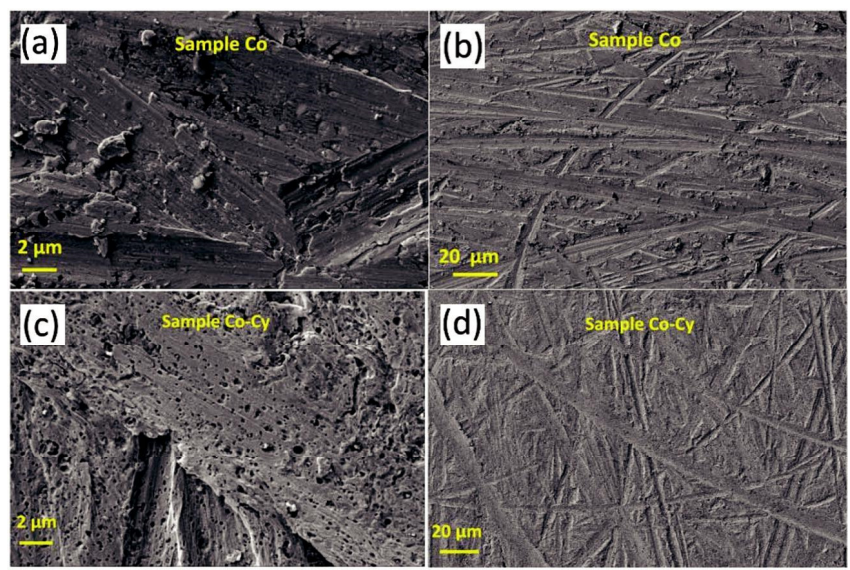

Fig. 6 SEM top view images of samples Co (a, b) and Co-Cy (c, d). Accelerating voltage: 5 kV; detector: secondary electron detector.

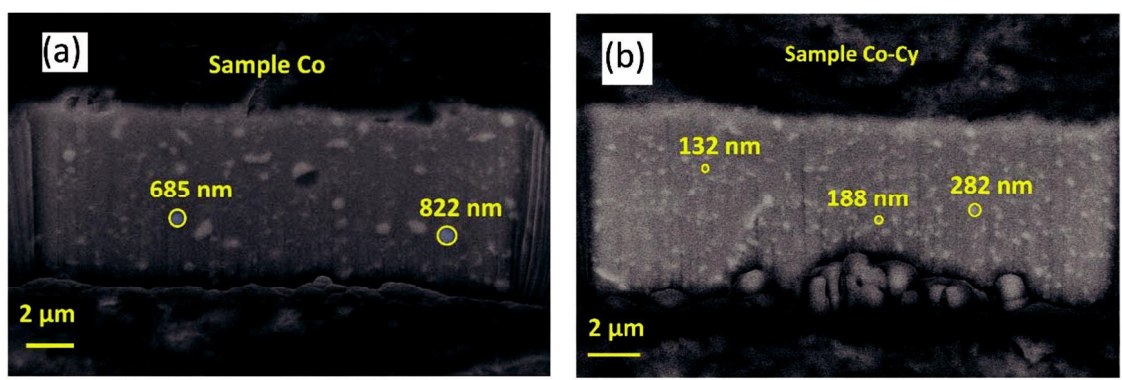

Fig. 7 Cross sectional analysis of untreated and surface modified steel derived from FIB-SEM experiments. The accelerating voltage was adjusted to 20 kV and the SEM images were acquired using a backscattered detector. Ga beam settings: 2 nA, 30 kV.

We speculate that the diffusion of ions caused by a momen-tum transfer $e^- \rightarrow M^+$ at high current densities[52] as discussed in one of our earlier contributions[20] is responsible for the changes of the surface composition as determined by XPS spectroscopy and summarized in Table S2.† XPS data of untreated X20CoCrWMo10-9 steel have been shown in our earlier contribution.[29] High resolution XPS spectra of sample Co-Cy were recorded after 4000 s of OER at 10 mA $cm^{-2}$ in pH 1 solution. The binding energies apparent in Fig. 8 suggest that no metallic Cr or Cr(VI) oxide is present.[53–55] Most likely, Cr is present in the form of Cr(III), either as $Cr_2O_3$, as Cr(III) hydroxide, or as an admixture of both species (Fig. 8). Iron in the form of FeOOH species clearly dominates at the surface of the surface-oxidized samples whereas only a very small signal located at ~706.7 eV can be assigned to metallic Fe (Fig. 8). To sum up, our XPS findings basically reveal that FeO(OH) is the driving force for OER on the surface of electro activated Co steel (sample Co-Cy).

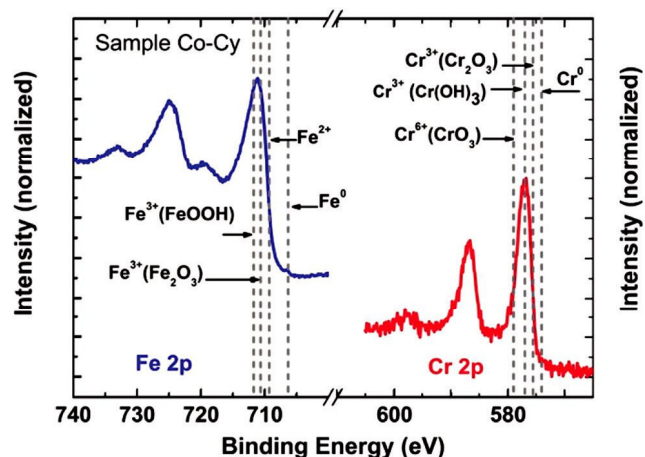

Fig. 8 High resolution XPS spectra of the sample Co-Cy. The binding energies of reference compounds are indicated by vertical lines as a guide to the eye. Left side: Fe 2p core level spectra. Right side: Cr 2p core level spectra.

## 4. Conclusions

All known electrocatalysts that durably and efficiently convert electricity at low pH values into oxygen, which is a requirement for their exploitation as electrodes in PEM electrolyzers, contain noble metals at least in the form of additives. This work evaluates the suitability of a cobalt containing steel for usage as an oxygen evolving electrode in acids with pH ≤ 1. Untreated X20CoCrWMo10-9 steel (sample Co) exhibited a reasonable current–voltage behavior under steady state con ditions. The potential required for 10 mA cm$^{-2}$ current density at pH 1 amounted to 1.924 V vs. RHE that corresponds to 696 mV overpotential. However, the OER performance of sample Co was found to be sensitive towards repeated dynamic variation of the voltage as seen by linear sweep vol-tammetry. In addition, the electrode exhibited a serious weight loss upon long term polarization in sulfuric acid (98.9 μg mm$^{-2}$). We showed that oxidation of X20CoCrWMo10-9 steel upon repetitive multicycling of the potential in 0.68 M LiOH leads to a reduction of the size of ferritic carbides, which is likely to be responsible for lower resistivity values as shown by EIS and results in better OER properties. Moreover, a cross sectional analysis of the outer zone of the catalyst did not reveal a classical substrate-layer architecture which to our experience also results in better electrocatalytic stability. The OER based overpotential amounted to 574 mV during long term chronopotentiometry performed at 10 mA cm$^{-2}$ current density in 0.05 M sulfuric acid. In addition to a substantially better OER based current–voltage relationship when compared to sample Co, the weight loss of the surface oxidized steel (sample Co-Cy) that occurred during long term usage was found to be reduced by around 60%. Notably, also, IrO$_2$–RuO$_2$ exhibited a mass loss upon usage as an oxygen evolving elec-trode. To the best of our knowledge, a similar activity and durability for OER in acids proved by comprehensive testing has not been shown for a catalyst solely consisting of cheap elements.

## Conflicts of interest
The authors declare no competing financial interest.

## Author contributions
H. S. had the idea to perform the experiments the manuscript is based on. He planned, performed and evaluated the electrochemical measurements and all sample preparations. H. S. wrote the draft of the manuscript. K. K. and J. W. planned, performed and evaluated the XPS measurements. K. M.-B. and J. S. planned and performed the BET measurements. D. D. evaluated the ICP OES measurements. W. H., M. S. and M. St. planned and per-formed the SEM experiments. U. K. planned and performed the FIB-SEM experiments. All authors have read and edited the manuscript.

## Acknowledgements
H. S., M. St., M. S. and W. H. were supported by the European Research Council (ERC-CoG-2014; project 646742 INCANA). The authors thank the German Research Foundation for funding a focused ion beam unit (INST 190/164-1 FUGG).

# References

1 F. Le Formal, W. S. Bouree, M. S. Prevot and K. Sivula, Chimia, 2015, 69(12), 789–798.
2 M. G. Walter, E. L. Warren, J. R. McKone, S. W. Boettcher, Q. Mi, E. A. Santori and N. S. Lewis, Chem. Rev., 2010, 110, 6446–6473.
3 T. R. Cook, D. K. Dogutan, S. Y. Reece, Y. Surendranath, T. S. Teets and D. G. Nocera, Chem. Rev., 2010, 11, 6474–6502.
4 A. J. Bard and M. A. Fox, Acc. Chem. Res., 1995, 28, 141–145.
5 J. Wang, H. Zhang and X. Wang, Small Methods, 2017, 1, 1700118.
6 L. Sun, L. Hammarstrom, B. Akermark and S. Styring, Chem. Soc. Rev., 2001, 30, 36–49.
7 H. Hu, Y. Fan and H. Liu, Int. J. Hydrogen Energy, 2009, 34, 8535–8542.
8 J. Tian, Y. Leng, Z. Zhao, Y. Xia, Y. Sang, P. Hao, J. Zhan, M. Li and H. Liu, Nano Energy, 2015, 11, 419–427.
9 L. L. Duan, F. Bozoglian, S. Mandal, B. Stewart, T. Privalov, A. LIobet and L. Sun, Nat. Chem., 2012, 4, 418–423.
10 A. Kudo and Y. Miseki, Chem. Soc. Rev., 2009, 38, 253–278.
11 P. Zhang, H. Chen, M. Wang, Y. Yang, J. Jiang, B. Zhang, L. Duan, Q. Daniel, F. Li and L. Sun, J. Mater. Chem. A, 2017, 5, 7564–7570.
12 Y. Matsumoto and E. Sato, Mater. Chem. Phys., 1986, 14, 397–426.
13 W. Zhou, X.-J. Wu, X. Cao, X. Huang, C. Tan, J. Tian, H. Liu, J. Wang and H. Zhang, Energy Environ. Sci., 2013, 6, 2921–2924.
14 M. W. Louie and A. T. Bell, J. Am. Chem. Soc., 2013, 135, 12329–12337.
15 L. Trotochaud, S. L. Young, J. K. Rannes and S. W. Boettcher, J. Am. Chem. Soc., 2014, 136, 6744–6753.
16 E. Umeshbabu, G. Rajeshkhanna, P. Justin and G. R. Rao, Solid State Electrochem., 2016, 20, 2725–2736.
17 F. Le Formal, N. Guijarro, W. S. Bourée, A. Gopakumar, M. S. Prévot, A. Daubry, L. Lombardo, C. Sornay, J. Voit, A. Magrez, P. J. Dyson and K. Sivula, Energy Environ. Sci., 2016, 9, 3448–3455.
18 K. S. Joya, Z. Ahmad, Y. F. Joya, A. T. Garcia-Esparza and H. J. M. de Groot, Nanoscale, 2016, 8, 15033–15040.
19 K. S. Joya, L. Sinatra, L. G. Abdul Halim, C. P. Joshi, M. N. Hedhili, O. M. Bakr and I. Hussain, Nanoscale, 2016, 8, 9695–9703.
20 H. Schäfer, D. M. Chevrier, P. Zhang, K. Kuepper, J. Stangl, K. M. Müller-Buschbaum, J. D. Hardege, J. Wollschlaeger, U. Krupp, S. Duehnen, M. Steinhart, L. Walder, S. Sadaf and M. Schmidt, Adv. Funct. Mater., 2016, 20(35), 6402–6417.
21 C. C. L. McCrory, S. Jung, J. C. Peters and T. F. Jaramillo, J. Am. Chem. Soc., 2013, 135(45), 16977–16987.
22 C. C. L. McCrory, S. Jung, I. M. Ferrer, S. M. Chatman and I. C. Peters, J. Am. Chem. Soc., 2015, 137, 4347–4357.
23 N. Danilovicet, R. Subbaraman, K.-C. Chang, S. H. Chang, Y. I. Kang, J. Snyder, A. P. Paulikas, D. Strmcnik, Y.-T. Kim, D. Myers, V. R. Stamenkovic and N. M. Markovic, J. Phys. Chem. Lett., 2014, 5, 2474–2478.
24 K. Sardar, E. Petrucco, C. I. Hiley, J. D. B. Sharman, P. P. Wells, A. E. Russell, R. J. Kashtiban, J. Sloan and R. Walton, Angew. Chem., Int. Ed., 2014, 53, 10960–10964.
25 T. Audichon, S. Morisset, T. W. Napporn, K. B. Kokoh, C. Comminges and C. Morais, ChemElectroChem, 2015, 2, 1128–1137.
26 O. Diaz-Morales, S. Raaijman, R. Kortlever, P. J. Kooyman, T. Wezendonk, J. Gascon, W. T. Fu and M. T. Koper, Nat. Commun., 2016, 7, 12363, DOI: 10.1038/ncomms12363.
27 N. Baumann, C. Cremers, K. Pinkwart and J. Tübke, Fuel Cells, 2017, 17, 259–267.
28 H. Schäfer, S. Sadaf, L. Walder, K. Kuepper, S. Dinklage, J. Wollschlaeger, L. Schneider, M. Steinhart, J. D. Hardege and D. Daum, Energy Environ. Sci., 2015, 8, 2685–2697.
29 H. Schäfer, D. M. Chevrier, K. Kuepper, P. Zhang, J. Wollschlaeger, D. Daum, M. Steinhart, C. Heß, U. Krupp, K. Müller-Buschbaum, J. Stangl and M. Schmidt, Energy Environ. Sci., 2016, 9, 2609–2622.
30 H. Schäfer, S. M. Beladi-Mousavi, L. Walder, J. Wollschläger, O. Kuschel, S. Ichilmann, S. Sadaf, M. Steinhart, K. Küpper and L. Schneider, ACS Catal., 2015, 5, 2671–2680.
31 H. Schäfer, K. Küpper, J. Wollschläger, N. Kashaev, J. D. Hardege, L. Walder, S. M. Beladi-Mousavi, B. Hartmann-Azanza, M. Steinhart, S. Sadaf and F. Dorn, ChemSusChem, 2015, 21(8), 3099–3110.
32 MetalRavne, 2390 Ravne na Koroškem, Slovenija, EU. http://www.metalravne.com.
33 A. Shinde, R. J. R. Jones, D. Guevarra, S. Mitrovic, N. Becera-Stasiewicz, J. A. Haber, J. Jin and J. M. Gregoire, Electrocatalysis, 2015, 6, 229–236.
34 A. Zadick, L. Dubau, U. B. Demirci and M. Chatenet, J. Electrochem. Soc., 2016, 163(8), F781–F787.
35 A. Zadick, L. Dubau, N. Sergent, G. Berthome and M. Chatenet, ACS Catal., 2015, 5, 4819–4824.
36 M. Huynh, D. K. Bediako and D. G. Nocera, J. Am. Chem. Soc., 2014, 136, 6002–6010.
37 M. M. Najafpour, K. C. Leonard, F.-R. F. Fan, M. M. A. Tabrizi, A. J. Bard, C. K. King'ondu, S. L. Suib, B. Haghighi and S. I. Allakhverdiev, Dalton Trans., 2013, 42, 5085–5091.
38 N. Suzuki, T. Horie, G. Kitahara, M. Murase, K. Shinozaki and Y. Morimoto, Electrocatalysis, 2016, 7, 115–120.
39 R. Frydendal, E. A. Paoli, I. Chorkendorff, J. Rossmeisl and I. E. L. Stephens, Adv. Energy Mater., 2015, 5, 1500991.
40 Y. Pi, Q. Shao, P. Wang, J. Guo and X. Huang, Adv. Funct. Mater., 2017, 27, 1700886.
41 M. Huynh, T. Ozel, C. Liu, E. C. Lau and D. G. Nocera, Chem. Sci., 2017, 8, 4779–4794.
42 P. P. Patel, M. K. Datta, O. I. Velikokhatnyi, R. Kuruba, K. Damodaran, P. Jampani, B. Gattu, P. M. Shanti, S. S. Damle and P. N. Kumta, Sci. Rep., 2016, 6, 28367, DOI: 10.1038/srep28367.

43 L. C. Seitz, C. F. Dickens, K. Nishio, Y. Hikita, J. Montoya, A. Doyle, C. Kirk, A. Vojvodic, H. Y. Hwang, J. K. Norskov and T. F. Jaramillo, Science, 2016, 353(6303), 1011–1014.
44 R. L. Doyle and M. E. G. Lyons, J. Solid State Electrochem., 2014, 18, 3271–3286.
45 I. J. Godwin, R. L. Doyle and M. E. G. Lyons, J. Electrochem. Soc., 2014, 161, F906–F917.
46 M. E. G. Lyons and S. Floquet, Phys. Chem. Chem. Phys., 2011, 13, 5314–5335.
47 E. Tsuji, A. Imanishi, K.-i. Fukui and Y. Nakato, Electrochim. Acta, 2011, 56, 2009–2016.
48 S. Cherevko, S. Geiger, O. Kasian, N. Kulyk, J.-P. Grote, A. Savan, B. RatnaShresta, S. Merzlikin, B. Breitbach, A. Luwig and K. J. J. Mayrhofer, Catal. Today, 2016, 262, 170–180.
49 C. Iwakura, K. Hirao and H. Tamura, Electrochim. Acta, 1977, 22, 335–340.
50 M. Godec, T. V. Pirtovšek, B. Š. Batič, P. McGuiness, J. Burja and B. Podgornik, Sci. Rep., 2015, 5, 16202, DOI: 10.1038/srep16202.
51 X. Yang, H. Li, A.-Y. Lu, S. Min, Z. Idriss, M. N. Hedhili, K.-W. Huang, H. Idriss and L.-J. Li, Nano Energy, 2016, 25, 42–50.
52 C. B. Lee, B. S. Kang, M. J. Lee, S. E. Ahn, G. Stefanovich, W. X. Xianyu, K. H. Kim, J. H. Hur, H. Yin, Y. Park, I. Yoo, J. B. Park and B. H. Park, Appl. Phys. Lett., 2007, 91, 082104.
53 M. C. Biesinger, B. P. Payne, A. P. Grosvenor, L. W. M. Lau, A. R. Gerson and R. St. C. Smart, Appl. Surf. Sci., 2011, 257, 2717–2730.
54 C. Klewe, M. Meinert, A. Boehnke, K. Kuepper, E. Arenholz, A. Gupta, J. M. Schmalhorst, T. Kuschel and G. Reiss, J. Appl. Phys., 2014, 115(12), 123903.
55 A. P. Grosvenor, B. A. Kobe, M. C. Biesinger and N. S. McIntrye, Surf. Interface Anal., 2004, 36, 1564–1574.